# Macroscopic proof of the Jarzynski-Wójcik fluctuation theorem for heat exchange

Yuki Sughiyama[1] and Sumiyoshi Abe[1,2*,3]

[1] Inspire Institute Inc., McLean, Virginia 22101, USA

[2] Department of Physical Engineering, Mie University, Mie 514-8507, Japan

[3] Institut Supérieur des Matériaux et Mécaniques Avancés, 44 F. A. Bartholdi,
    72000 Le Mans, France

**Abstract.** In a recent work, Jarzynski and Wójcik (2004 *Phys. Rev. Lett.* **92** 230602) have shown by using the properties of Hamiltonian dynamics and a statistical mechanical consideration that, through contact, heat exchange between two systems initially prepared at different temperatures obeys a fluctuation theorem. Here, another proof is presented, in which only macroscopic thermodynamic quantities are employed. The detailed balance condition is found to play an essential role. As a result, the theorem is found to hold under very general conditions.

___________________________

∗ Permanent address.

## 1. Introduction

Consider two finite systems, *A* and *B*, initially prepared in equilibrium states at different temperatures, $T_A$ and $T_B$ (say, $T_A > T_B$), separately, contact them at time $t = 0$ and separate them again at $t = \tau$. Then, how much heat is transferred from *A* to *B*? In a recent work [1], Jarzynski and Wójcik have posed this question and proved that the distribution, $P_\tau(\Delta Q)$, of the net heat transfer $\Delta Q$ obeys the fluctuation theorem [2-4] of the following form:

$$\frac{P_\tau(\Delta Q)}{P_\tau(-\Delta Q)} = \exp\left[\left(\frac{1}{T_B} - \frac{1}{T_A}\right)\Delta Q\right], \tag{1}$$

which shows how the heat can flow from a cold body to a hot one with a finite value of probability. To derive this relation, the authors of Ref. [1] discuss time reversal of Hamiltonian dynamics and assume that $T_A$ and $T_B$ remain constant during the time interval $[0, \tau]$, that is, each equilibrium state represented by the canonical ensemble does not change.

In this paper, we present another proof of Eq. (1) based on an approach in the style of the Onsager-Machlup theory [5]. We develop our discussion by making use of *macroscopic thermodynamic quantities of three systems*, *A*, *B and E*. An advantageous point of this approach is that the constancy of the temperatures does not have to be assumed. In addition, one of the systems, *B*, can be small. Thus, Eq. (1) is shown to hold under very general conditions. It is our opinion that this independent proof casts new light on the issue.

## 2. Onsager-Machlup theory and detailed balance condition on thermodynamic variables

Let us start our discussion with recalling the Onsager-Machlup theory of fluctuations in irreversible processes. Consider the objective system surrounded by the environment and assume that they are initially not in equilibrium. The internal energy of the objective system, $\phi$, may evolve in time along a relaxation process according to the Langevin equation [5]: $d\phi/dt = L\, dS^{tot}/d\phi + \xi$, where $L$, $S^{tot}$ and $\xi$ are the transport coefficient, the total entropy (of the objective and environmental systems) and a noise, respectively. Near equilibrium, the total entropy is well approximated by a quadratic function [5-7]:

$$S^{tot}(\phi) = \text{const.} - \frac{1}{2}\alpha\phi^2 \qquad (2)$$

with $\alpha > 0$, provided that $\phi_0$ yielding $S^{tot}(\phi_0) = \max$ is taken to be zero for the sake of simplicity, that is, $\phi$ is the variable describing fluctuation of the energy. Therefore, the thermodynamic force, $dS^{tot}/d\phi$, is linear, and the Langevin equation becomes

$$\frac{d\phi}{dt} = -\lambda\phi + \xi, \qquad (3)$$

where $\lambda = L\alpha$ is a positive constant. Following Onsager and Machlup, we require the noise to be the unbiased Gaussian white noise:

$$\overline{\xi(t)} = 0, \qquad \overline{\xi(t)\,\xi(t')} = 2D\delta(t-t'), \tag{4}$$

where the over-bar stands for the average over the noise distribution and $D$ is the diffusion constant. Take a time interval $[t_1, t_2]$ and impose the conditions, $\phi(t_1) = X$ and $\phi(t_2) = Y$. The transition probability from $X$ to $Y$ is given by the following functional integral [5]:

$$f(Y, t_2 \mid X, t_1) = N \int_{\phi(t_1)=X}^{\phi(t_2)=Y} D\phi \, \exp\left(-\int_{t_1}^{t_2} dt\, \pounds\right), \tag{5}$$

where $N$ is a normalization factor, and

$$\pounds = \frac{1}{4D}\left(\frac{d\phi}{dt} + \lambda\phi\right)^2 \tag{6}$$

is the "thermodynamic Lagrangian".

Now, let us consider time reversal: $t = -\tilde{t}$ [8,9]. Under this operation, $\phi$ is assumed to transform as a scalar variable, i.e., $\tilde{\phi}(\tilde{t}) = \phi(t)$. Accordingly, the thermodynamic Langragian transforms as follows:

$$\pounds\big(\phi(t), d\phi(t)/dt\big) = \pounds\big(\tilde{\phi}(\tilde{t}), d\tilde{\phi}(\tilde{t})/d\tilde{t}\big) + \frac{dS^{\text{tot}}(\tilde{\phi}(\tilde{t}))}{d\tilde{t}}. \tag{7}$$

Upon deriving this equation, we have used Eq. (2) as well as the fluctuation dissipation theorem, $D = L$. Therefore, the transition probability changes as

$$f(Y, t_2 | X, t_1) = N \, e^{S^{tot}(Y) - S^{tot}(X)} \int_{\tilde{\phi}(-t_1)=X}^{\tilde{\phi}(-t_2)=Y} D\tilde{\phi} \exp\left[-\int_{-t_2}^{-t_1} d\tilde{t} \, \pounds\left(\tilde{\phi}(\tilde{t}), d\tilde{\phi}(\tilde{t})/d\tilde{t}\right)\right]. \quad (8)$$

Finally, doing the shift, $\hat{t} = \tilde{t} + t_1 + t_2$, and noticing $\tilde{\phi}(\tilde{t}) = \hat{\phi}(\hat{t})$ as well as the invariance of the functional integral part under time translation, we obtain the detailed balance condition:

$$f(Y, t_2 | X, t_1) \rho_\infty(X) = f(X, t_2 | Y, t_1) \rho_\infty(Y), \quad (9)$$

where we have used Einstein's relation [6,7] for the distribution of fluctuations around equilibrium, $\rho_\infty(\phi) \propto \exp[S^{tot}(\phi)]$, with Boltzmann's constant being set equal to unity.

The detailed balance condition is usually thought of as a remnant of microscopic reversibility [10]. It should however be noticed that the quantities treated here are the macroscopic thermodynamic variables. We shall see how this detailed balance condition plays an essential role in proving Eq. (1).

### 3. Macroscopic proof of fluctuation theorem for heat exchange

We are in a position to present another proof of the Jarzynski-Wójcik theorem. The physical setup of our system is as follows. Here, system $A$ is assumed to be much larger than system $B$. Initially, $A$ is in an equilibrium state with temperature $T_A$, whereas $B$ is in a relaxed state in equilibrium with its surrounding environment $E$ with temperature, $T_E \equiv T_B (< T_A)$. That is, $A$ is initially separated from the composite $B + E$ system. Considering the third system, $E$, is in marked contrast to the setup in Ref. [1]. The probability of finding B in the state with $\phi$ is given by [6,7]

$$\rho_{\text{initial}}^{B}(\phi) \propto \exp\left[S^{B}(\phi) + S^{E}(\phi^{B+E} - \phi)\right], \tag{10}$$

where $S^{B}$ ($S^{E}$) and $\phi^{B+E} \equiv \phi + \phi^{E}$ are the entropy of B (E) and the total energy of $B+E$, respectively. (As usual, the interaction between B and E is weak and its energy is assumed to be negligible.)

Now, separate B from E and bring it into contact with A at $t = 0$. Then, separate B from A at $t = \tau$. During the time interval $[0, \tau]$, the relaxation process is described by the Onsager-Machlup theory. Setting $t_1 = 0$ and $t_2 = \tau$, the detailed balance condition in Eq. (9) becomes

$$f(Y, \tau | X, 0)\, \rho_{\infty}^{B}(X) = f(X, \tau | Y, 0)\, \rho_{\infty}^{B}(Y). \tag{11}$$

In this equation, $\rho_{\infty}^{B}(\phi)$ denotes the state of B sufficiently relaxed in A, that is,

$$\rho_{\infty}^{B}(\phi) \propto \exp\left[S^{A}(\phi^{A+B} - \phi) + S^{B}(\phi)\right], \tag{12}$$

where $S^{A}$ and $\phi^{A+B} \equiv \phi^{A} + \phi$ are the entropy of A and the total energy of $A+B$, respectively. Here, we have assumed that the functional form (not the value) of $S^{B}(\phi)$ does not change when B is separated from E and when its contact with A is made.

Let us calculate the probability distribution, $P_{\tau}(\Delta Q)$, that the net heat transfer from A to B during $[0, \tau]$ is $\Delta Q$:

$$P_\tau(\Delta Q) = \langle \delta(\Delta Q - (Y - X)) \rangle$$

$$= \iint dX\, dY\, \delta(\Delta Q - (Y - X))\, f(Y, \tau | X, 0)\, \rho^B_{\text{initial}}(X). \tag{13}$$

With the help of Eq. (10), this equation is rewritten as

$$P_\tau(\Delta Q) = N_0 \iint dX\, dY\, \delta(\Delta Q - (Y - X))\, f(Y, \tau | X, 0)\, \rho^B_\infty(X)$$
$$\times \exp\left[S^E(\phi^{B+E} - X) - S^A(\phi^{A+B} - X)\right], \tag{14}$$

where $N_0$ is a normalization constant. Furthermore, from the detailed balance condition in Eq. (11), it follows that

$$P_\tau(\Delta Q) = N_0 \iint dX\, dY\, \delta(\Delta Q - (Y - X))\, f(X, \tau | Y, 0)\, \rho^B_\infty(Y)$$
$$\times \exp\left[S^E(\phi^{B+E} - X) - S^A(\phi^{A+B} - X)\right]$$

$$= \iint dX\, dY\, \delta(\Delta Q - (Y - X))\, f(X, \tau | Y, 0)\, \rho^B_{\text{initial}}(Y)$$
$$\times \exp\left[S^A(\phi^{A+B} - Y) - S^A(\phi^{A+B} - X) + S^E(\phi^{B+E} - X) - S^E(\phi^{B+E} - Y)\right]. \tag{15}$$

Since $\phi^A, \phi^E \gg \phi$, the first-order approximation leads to

$$S^A(\phi^{A+B} - Z) = S^A(\phi^{A+B}) - \frac{1}{T_A}(\phi^{A+B} - Z), \tag{16}$$

$$S^E(\phi^{B+E} - Z) = S^E(\phi^{B+E}) - \frac{1}{T_E}(\phi^{B+E} - Z) \tag{17}$$

with $Z = X, Y$, where $1/T_A = \partial S^A / \partial \phi^A$ and so on. Using Eqs. (16) and (17) in Eq. (15) and interchanging the integration variables, we finally obtain

$$P_\tau(\Delta Q) = \iint dX\, dY\, \delta(\Delta Q - (Y - X))\, f(X, \tau | Y, 0)\, \rho_{\text{initial}}^B(Y)$$

$$\times \exp\left[\left(\frac{1}{T_E} - \frac{1}{T_A}\right)(Y - X)\right]$$

$$= \exp\left[\left(\frac{1}{T_E} - \frac{1}{T_A}\right)\Delta Q\right] \iint dX\, dY\, \delta(-\Delta Q - (Y - X))\, f(Y, \tau | X, 0)\, \rho_{\text{initial}}^B(X)$$

$$= \exp\left[\left(\frac{1}{T_E} - \frac{1}{T_A}\right)\Delta Q\right] \langle \delta(-\Delta Q - (Y - X)) \rangle$$

$$= \exp\left[\left(\frac{1}{T_E} - \frac{1}{T_A}\right)\Delta Q\right] P_\tau(-\Delta Q). \tag{18}$$

Recalling the fact that $T_E$ is identical to the initial temperature of $B$, $T_B$, one sees that Eq. (18) proves Eq. (1).

## 4. Conclusion

We have presented another proof of the fluctuation theorem for heat exchange based on the Onsager-Machlup macroscopic theory for fluctuations in irreversible processes. We have shown that the theorem holds under the conditions more general than those in the work of Jarzynski and Wójcik. We have seen how the detailed balance condition plays an essential role in the proof.

**Acknowledgments**

The authors would like to thank Professor Christopher Jarzynski for discussions. The work of S. A. was supported in part by Grant-in-Aid for Scientific Research of the Japan Society for the Promotion of Science.

**References**

[1] Jarzynski C and Wójcik D K, 2004 *Phys. Rev. Lett.* **92** 230602

[2] Evans D J, Cohen E G D and Morriss G P, 1993 *Phys. Rev. Lett.* **71** 2401; (E) **71** 3616

[3] Gallavotti G and Cohen E G D, 1995 *Phys. Rev. Lett.* **74** 2694

[4] Gallavotti G and Cohen E G D, 1995 *J. Stat. Phys.* **80** 931

[5] Onsager L and Machlup S, 1953 *Phys. Rev.* **91** 1505

[6] Einstein A, 1910 *Ann. der Phys.* 33 1275

[7] Landau L D and Lifshitz E M, 1980 *Statistical Physics* 3rd edn
    (Pergamon Press: Oxford)

[8] Taniguchi T and Cohen E G D, 2006 *J. Stat. Phys.* **126** 1

[9] Sughiyama Y and Abe S, e-print arXiv:0803.1429

[10] Tolman R C, 1979 The Principles of Statistical Mechanics (Dover: New York)